\documentclass[11pt,aps,prd,preprint,superscriptaddress]{revtex4}
\usepackage{amsmath}
\usepackage[dvips]{graphicx}
\usepackage[english]{babel}
\newcommand{\be}{\begin{equation}}
\newcommand{\ee}{\end{equation}}
\newcommand{\ben}{\begin{eqnarray}}
\newcommand{\een}{\end{eqnarray}}

\begin{document}
\title{Does the entropy of the Universe tend to a maximum?}
\author{Diego Pav\'{o}n\footnote{E-mail: diego.pavon@uab.es} and
Ninfa Radicella\footnote{E-mail: ninfa.radicella@uab.cat}}
\affiliation{Departamento de F\'{\i}sica, Universidad Aut\'{o}noma
de Barcelona, 08193 Bellaterra (Barcelona), Spain.}
\begin{abstract}
Ordinary, macroscopic systems, naturally tend to a state of
maximum entropy compatible with their constraints. However, this
might not hold for gravity-dominated systems since their entropy
may increase without bound unless this is precluded by the
formation of a black hole. In this short note we suggest, based on
the Hubble expansion history,  that our Universe likely behaves as
an ordinary system, i.e., that its entropy seems to tend to some
maximum value.
\end{abstract}
\maketitle

\noindent As is well known, systems dominated by electromagnetic
forces spontaneously tend to some equilibrium state compatible
with the constraints the system is subjected to. This constitutes
the hard core of the empirical basis of the second law of
thermodynamics. According to the latter isolated, macroscopic
systems, evolve to the maximum entropy state consistent with their
constraints \cite{callen}. This implies two separate things:
first, the entropy, $S$,  of isolated systems cannot decrease,
i.e., $S' \geq 0$, where the prime means derivative with respect
to the relevant, appropriate variable. Secondly, it must
be a convex function of the  said variable, $S'' < 0$,
at least at the last stage of the evolution.
\  \\

\noindent However, this is not necessarily true when gravity plays
a significant role. The entropy of systems dominated by gravity
must still increase (as formulated by the generalized second law
of thermodynamics \cite{jakob1,jakob2,sewell}), but -at least in
Newtonian gravity-  it may grow unbounded. That is to say, while
the relationship $S' > 0$ remains in place it may well happen that
$S'' > 0$ when the variable approaches its final value; in such a
case, no maximum entropy state would be achievable. This is
illustrated by the gravothermal catastrophe; namely, the final
stage of $N$ gravitating point masses enclosed in a perfectly
reflecting, rigid, sphere whose radius exceeds some critical value
\cite{antonov,lyndenbell68}.
\   \\

\noindent Nevertheless, when Newtonian gravity is replaced by
general relativity  a black hole is expected to form at the center
of the sphere whereby the said catastrophe is prevented and the
entropy does not diverge. Though the black hole will tend to
evaporate, which will again increase the total entropy, it will
likely arrive to an equilibrium state with its own radiation -see
e.g. \cite{kip}- whence the end stage of the whole process will be
characterized by a state of maximum, finite,  entropy.
\  \\

\noindent At any rate, as far as we know, the possibility of
realistic processes in which no equilibrium state is achievable,
because $S'' > 0 $ in the long run, cannot be ruled out right
away. Then, the question arises whether the entropy of the
Universe mimics the first set of systems (ordinary systems) or the
second -exotic- set. The aim of this short note is to shed light
on this issue. As we shall see, our analysis suggests that it
mimics the entropy of systems falling in the first set, i.e., in
this respect, the Universe behaves as an ordinary system. To make
matters simpler we will work in the framework of general relativity
and assume the Universe sufficiently well
described at large scales by the spatially-flat
Friedmann-Robertson-Walker (FRW) metric of scale factor $a$.
\  \\

\noindent We shall base our study, on the one hand, on the close
connection between entropy and area of cosmic horizons and, on the
other hand, on observational results about the history of the
Hubble factor, $H = \dot{a}/a$, of the FRW metric. While these
results are still rather preliminary, it seems beyond doubt that
$H$ decreases with expansion, i.e., $H'(a)< 0$, where the prime
means derivative with respect to $a$, and that $H''(a) > 0$. This
is fully consistent with recent studies on the impact of
hypothetical transient periods of acceleration-deceleration on the
matter growth \cite{prd_linder} and on the radiation power
spectrum \cite{jcap_linder} from the decoupling era, $a \simeq
10^{-5}$, to $a = 0.5$, and with the study of Serra {\it et al.}
\cite{serra} that shows that equation of state parameter of dark
energy has not been noticeably  varied between $a=0.5$ and $1$,
where the normalization $a_0=1$ is understood.  These studies
strongly suggest the absence of the said hypothetical periods. A
further and crucial observation is that the cosmic expansion is
accelerating at present \cite{riess,perlmutter,komatsu,amanullah},
thereby the current value of the deceleration parameter results
negative, $q_{0} = -[1\, + (aH'/H)]_{0} < 0 $.
\ \\

\noindent Nowadays it is widely accepted  that, sooner or later,
the entropy of the expanding Universe is to be dominated by the
entropy of the horizon, which is proportional to the area of the
latter, i.e., $S \propto {\cal A}$. This relation is valid, at
least, in Einstein gravity \cite{prd_gary}.  We will deal here
with apparent horizons that are endowed with thermodynamical
properties, formally identical to those of event horizons
\cite{cai1,cai2}. The apparent horizon is defined as the
marginally trapped surface with vanishing expansion of radius
$\tilde{r}=a(t) r$ \cite{cqg_bak}. In the simplest case of a
spatially flat FRW universe $\tilde{r}_A=H^{-1}$ and
\be {\cal A} \propto H^{-2} \, . \label{area1} \ee
\noindent Assuming that the Universe expands for ever (i.e., $H >
0$ at all times) and behaves from the thermodynamical standpoint
as any other macroscopic system, it is natural to expect that it
approaches equilibrium, characterized by a state of maximum
entropy compatible with the constraints. This translates into the
inequalities, ${\cal A}' > 0$ at all times and ${\cal A}'' < 0$ at
least as $a \rightarrow \infty$.
\  \\

From (\ref{area1}) we have
\be {\cal A}' \propto -\frac{2 \, H'}{H^{3}} \, , \qquad {\rm and}
\qquad {\cal A}'' \propto \frac{2}{H^{2}} \, \left[3 \,
\left(\frac{H'}{H}\right)^{2} \, - \, \frac{H''}{H}\right] \, .
\label{aderivatives} \ee
\noindent Measurements of the  Hubble factor at different
redshifts \cite{simon,stern} plus numerical simulations
\cite{crawford,carvalho-alcaniz} fairly suggest that $H'(a) < 0$
and $H''(a) > 0$, at least in the interval $0.4 \leq a \leq 1$
-see Figs. 1(c) and 3(a) in \cite{carvalho-alcaniz}, and Fig.
\ref{fig:H(a)} below. The latter shows the projected evolution of
the Hubble function in terms of the scale factor in the said
interval; the set of points was adapted from Fig. 3(a) of
\cite{carvalho-alcaniz}, which results from numerical simulations
-assuming a precision of  1\%- of H(z) measurements from luminous
red galaxies \cite{crawford}, plus the recently measured value of
the Hubble constant, $H_{0} = 74.2 \pm 3$ Km/s/Mpc -see Riess {\it
et al.} \cite{apj_Riess}.
\  \\

\noindent The overall behavior these figures show is shared by the
spatially flat $\Lambda$CDM model which seems to pass fairly well
most, if not all, observational tests. This implies that whatever
the ``right" cosmological model turns out to be, it will not
substantially differ, observationally, from the $\Lambda$CDM.
Since there is no apparent reason for this trend to change in the
future (it would if the Universe expansion were dominated by
phantom dark energy) we shall assume that the inequalities of
above will stay in place also for $a > 1$. In consequence ${\cal
A}'$ will result positive-definite, however $ {\cal A}''$ may bear
any sign. By imposing that ${\cal A}''$ should be negative, the
constraint
\be 3 \, \left(\frac{H'}{H}\right)^{2} \, < \, \frac{H''}{H}
\label{Hconstraint} \ee
readily follows. \\

\noindent From the set of conditions $H' < 0$,  $H''> 0$, and $q <
0$ -the latter holding only from some ``recent time" on-, it can
be demonstrated that for sufficiently large scale factor onwards
the inequality (\ref{Hconstraint}) is to be satisfied and,
accordingly, ${\cal A}'' < 0 $. Effectively, bear in mind that $q
= -[1+(aH'/H)]$; then $ H'/H = -(1+q)/a$. Since $H'$ and $q$ are
negative  the numerator of last expression stays bounded (it lies
in the range $0 \leq 1+q \leq 1$), whence the left hand side of
(\ref{Hconstraint}) vanishes in the long run.
\  \\

\noindent Inspection of panels (c) of Fig. 1 and (d) of Fig. 3 in
\cite{carvalho-alcaniz}, as well as Fig. \ref{fig:H(a)} below,
suggests that the data points can be roughly approximated by the
simple expressions
\begin{equation}
H = H_{*} \, \exp{(\lambda/a)} \, \qquad  {\rm and} \qquad H =
H_{*} \, (1 \, + \, \lambda \, a^{-n}) \, , \label{2H(a)}
\end{equation}
where $H_{*} = H (a \rightarrow \infty) > 0$, $\lambda > 0$, and
$n > 1$. Both functions describe ever expanding universes with $H'
< 0$ and $H''> 0$. By inserting the first one in
(\ref{Hconstraint}) one obtains that ${\cal A}'' < 0$ from the
instant the Universe starts accelerating onwards, namely, for $a
\geq \lambda$. By fitting (\ref{2H(a)}.1) to the set of points
displayed in Fig. \ref{fig:H(a)} (dashed line) we find that $H_{*}
= 42.6 \pm 0.4$ km/s/Mpc and $\lambda = 0.550 \pm 0.005$, both at
95\% confidence level (CL).
\  \\

\noindent For the cosmic expansion  described by (\ref{2H(a)}.2)
the transition from deceleration to acceleration occurs when
$a_{tr} = \left[\lambda (n-1) \right]^{1/n}$, and ${\cal A}'' < 0$
for $a
> \left[\lambda \, \left(\frac{3n}{n+1} \, - \, 1\right)
\right]^{1/n}$. The best fit values, at 95\% CL, of the parameters
to the set of points in Fig. \ref{fig:H(a)} are $H_{*} = 54.78 \pm
0.06$ km/s/Mpc, $\lambda = 0.3535 \pm 0.0013$ and $n = 1.928 \pm
0.002$ -see the dot-dashed line  in the said figure.  For
completeness and comparison we have also drawn the curve
corresponding to spatially flat $\Lambda$CDM model, $ H(a) = H_{0}
\, \sqrt{\Omega_{m0} \, a^{-3} \, + \, (1 \, - \,\Omega_{m0})}$
with $H_{0} = 73.3$ Km/s/Mpc and $\Omega_{m0} = 0.29$ which follow
from the fit to the set of points (solid line).

\begin{figure}[!hb]
  \begin{center}
    \begin{tabular}{c}
       \resizebox{120mm}{!}{\includegraphics{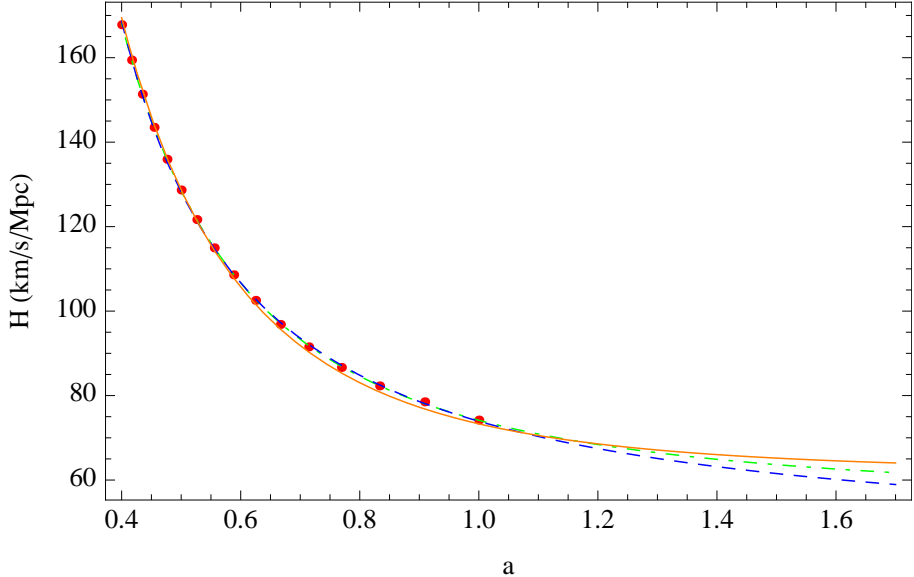}} \\
    \end{tabular}
    \caption{The string of points show the Hubble history in the interval $0.4 \leq a \leq 1$.
    The one  at $a =1$ indicates the Hubble constant value, $H_{0}$, as measured by Riess {\it et al.}
    \cite{apj_Riess}. The other fourteen points correspond to simulated values of the
    Hubble function assuming an accuracy of 1\% in the $H(a)$ observations according to Carvalho
    and Alcaniz (Fig. 3 (a) in Ref. \cite{carvalho-alcaniz}). The dashed, dot-dashed, and solid lines
    are the best fit curves of the models represented by Eq. (\ref{2H(a)}.1),
    Eq. (\ref{2H(a)}.2), and the spatially flat $\Lambda$CDM
    model, respectively.}
    \label{fig:H(a)}
  \end{center}
\end{figure}
\  \\

\begin{figure}[!hb]
  \begin{center}
    \begin{tabular}{c}
       \resizebox{120mm}{!}{\includegraphics{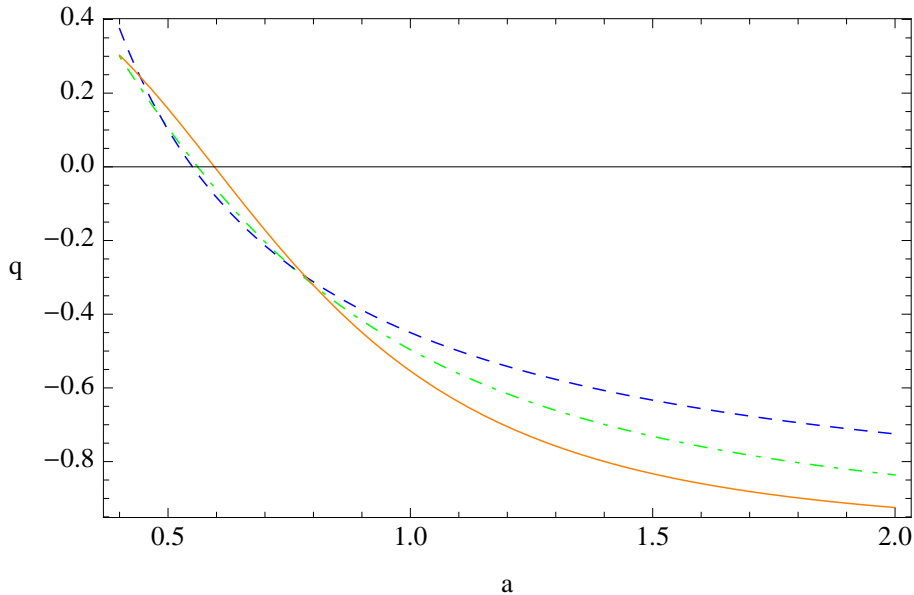}} \\
    \end{tabular}
    \caption{The deceleration parameter, $q = -(1 \, + \, aH'/H)$,  as a function of the scale
    factor. The dashed, dot-dashed, and solid lines correspond to the models of Eq. (\ref{2H(a)}.1),
    Eq. (\ref{2H(a)}.2), and the spatially flat $\Lambda$CDM model, respectively.}
    \label{fig:q(a)}
  \end{center}
\end{figure}

\noindent For completeness, Fig. \ref{fig:q(a)} presents the
evolution of deceleration parameter of the models considered.
Though not shown, in all the cases $q(a \rightarrow \infty) = -1.$
\  \\

\noindent At this point it is sobering to note that not all Hubble
functions that fulfill the observational restrictions $H' < 0$ and
$H''> 0$ comply with the inequality (\ref{Hconstraint}). This is,
for instance, the case of the expansion laws $H = H_{*} \,
[\exp{(\lambda \, a^{-1})} \, - \, 1] $ and $H = H_{*} \,
\exp{(-\lambda \, a)}$ (with $\lambda > 0$). Clearly, the entropy
of a universe that obeyed any of these two laws would increase
without bound in the long run, similarly to the entropy of
Antonov's sphere in Newtonian gravity. Note, however, that the
said functions do not correspond to realistic universes. In the
first case the universe never accelerates; in the second one the
universe accelerates at early times (when $a < 1/\lambda$) to
decelerate for ever afterwards. Thus, both of them are grossly
incompatible with observations. This suggests that Hubble
functions that satisfy the constraints $H' < 0$ and $H'' > 0$ but
violate Eq. (3) (thereby leading an unbound entropy as $a
\rightarrow \infty$) should be dismissed.
\   \\

\noindent Finally, as it can be checked, Eq. (3) would be
satisfied by quintessence-like dark energy models but not by
phantom dominated models \cite{radicella11a, radicella11b}.
Moreover, dynamical models that tend to de Sitter expansion at
late times approach equilibrium since the entropy function shows
an horizontal asymptote, a maximum entropy equilibrium state,
towards which the system tends from below.

\noindent We have not considered the entropy of matter and/or
fields driving the expansion of the Universe. The reasons for this
are as follows. $(i)$ As said above, except for the case of
phantom fields, these are subdominant in the long run whence, in
any case, the entropy of the horizon will eventually prevail.
Things are, nonetheless, different in the case of phantom fields.
They present the feature $S''> 0$ and could invalidate our
argument if they drove the accelerated expansion. However, as is
well known, they suffer from inherent quantum instabilities
\cite{prd_carroll,prd_cline} whereby they can hardly be considered
serious candidates for dark energy; this is why we ignored them.
$(ii)$ The entropy generated at small scales by matter and or
radiation through a variety of dissipative processes obeys the
second law of thermodynamics; taking it into account would only
strengthen our argument.
\  \\

\noindent Altogether, we may tentatively conclude that the entropy
of the Universe -like that of any ordinary system-, rather than to
increase indefinitely (as is the case of  Antonov's sphere),
appears to tend to some maximum value, possibly of the order of
$H^{-2}$ when $a \rightarrow \infty$. To reach a firmer conclusion
on this issue more abundant and accurate measurements regarding
the Hubble history of the Universe are needed.

\acknowledgments{We are grateful to  Fernando Atrio-Barandela and
Jos\'{e} Gaite for constructive remarks. NR is funded by the
Spanish Ministry of Education through the ``Subprograma Estancias
de J\'{o}venes Doctores Extranjeros, Modalidad B", Ref:
SB2009-0056. This work was partly supported by the Spanish
Ministry of Science and Innovation under Grant
FIS2009-13370-C02-01, and the ``Direcci\'{o} de Recerca de la
Generalitat" under Grant 2009SGR-00164.}


\end{document}